\documentstyle[12pt]{article}
\def\be{\begin{equation}}
\def\ee{\end{equation}}
\def\realnumb{{\rm I\!R}}
\newcommand{\lsim}{\mbox{\raisebox{-.3em}{$\stackrel{<}{\sim}$}}}

\begin{document}

\title{Strategies to measure a quantum state}
\author{Franz Embacher${}^\dagger$ and Heide Narnhofer${}^{*}$\\ \\
        Institute for Theoretical Physics\\
        University of Vienna\\
        Boltzmanngasse 5, A-1090 Wien, Austria\\
        \\
        }
\date{}
\maketitle

\begin{abstract}
We consider the problem of determining the mixed quantum state
of a large but finite number of identically prepared quantum
systems from data obtained in a sequence of ideal (von Neumann)
measurements, each performed on an individual copy of the system.
In contrast to previous approaches, we do not average over the
possible unknown states but work out a ``typical'' probability
distribution on the set of states, as implied by the experimental
data. As a consequence, any measure of knowledge about the unknown
state and thus any notion of ``best strategy'' (i.e. the choice of
observables to be measured, and the number of times they are
measured) depend on the unknown state. By learning from previously
obtained data, the experimentalist re-adjusts the observable to be
measured in the next step, eventually approaching an optimal
strategy.
\bigskip

We consider two measures of knowledge and exhibit all ``best'' strategies for
the case of a two-dimensional Hilbert space. Finally, we discuss some features
of the problem in higher dimensions and in the infinite dimensional case.
\bigskip

\noindent {\it PACS-numbers}: 03.65.-w, 03.65.Ta.
\end{abstract}

\noindent \\
{\small ${}^\dagger$ E-mail: fe@ap.univie.ac.at\\
${}^{*}$ E-mail: narnh@ap.univie.ac.at}


\section{Introduction}
\setcounter{equation}{0}

The topic of this paper is the problem of determining a quantum state from measurement data.
We consider a quantum system described by a Hilbert space ${\cal H}$ of
(finite) dimension $d$.
Given a large but finite number of copies of the system, all prepared in the same
quantum state $\tau$, we shall be allowed to perform an arbitrary (ideal) measurement
in each copy.
What knowledge about the state $\tau$ do we have after these measurements,
and what is the best strategy to maximize the information gained?
Several authors have considered problems of this type. Their approaches differ in
some respects, in particular regarding the measurement strategy and
the way how the knowledge about the unknown state $\tau$ is quantified.
\bigskip

The strategy analyzed by Wootters and Fields
\cite{WoottersFields1989} consists of choosing -- once and for all
-- a family of $d+1$ observables, and measuring each one in a
separate copy of the system an equal number of times. The
knowledge gained in these measurements is quantified in terms of
the average (over all possible unknown states) of an appropriately
defined ``uncertainty volume'' in the set of states (which
essentially stems from the Shannon \cite{Shannon1948} information
measure). They arrive at the result that the average gain
of knowledge in such a procedure is maximal if the $d+1$
observables measured are mutually unbiased (complementary). This
optimal strategy is, by definition, independent of the actual
(unknown) state $\tau$. Their paper is sometimes referred to as
proving that the use of mutually unbiased observables is the most
efficient determination of an unknown quantum state by means of
successive measurements.
\bigskip

Peres and Wootters \cite{PeresWootters1991} conjectured that an appropriately
designed single combined measurement on a number of identically prepared copies
of a quantum system is more efficient than a sequence of measurements on the
individual systems (a sequential measurement). Moreover, they provided evidence that generalized
(POVM based, see Refs. \cite{Helstrom1976}\cite{Peres1990}\cite{Peres1995})
measurements are more effective than ideal measurements of the von Neumann type \cite{vonNeumann1955}.
Their measure of knowledge is based on the Shannon information measure as well.
\bigskip

In a special scenario, Massar and Popescu \cite{MassarPopescu1995}
showed that a combined measurement is more efficient than a
sequential one, thus proving (``not in its letter, but in its
spirit'') Peres and Wootters' conjecture. In their work, knowledge
is measured by a ``score'' function defined as the average (over
all possible unknown states) of an expression quantifying the
difference between a candidate state and the unknown one.
\bigskip

In apparent contradiction to these results, Brody and Meister
\cite{BrodyMeister1996} showed that the minimum Bayesian decision
cost -- taking properly into account what is known {\it \`{a} priori}
about the unknown state -- is the same for sequential as for
combined measurements. They pointed out that the optimal strategy
in determining a quantum state depends on the details of the
approach, in particular on how the {\it \`{a} priori} knowledge is
treated.
\bigskip

In order to help clarifying these issues, we present a further
approach to the state determination problem. Thereby, we focus on
the original scenario of a sequence of ideal (von Neumann)
measurements on individual copies of the system. We first
compute a ``typical'' probability distribution on the set ${\cal
S}$ of states achieved after a (large) number of measurements,
thereby {\it retaining the dependence on the unknown state} $\tau$
throughout the analysis. In other words: we will {\it not} perform
an average over all possible unknown states. Thus, any measure of
knowledge (of which we discuss two variants, one being related to
the ``uncertainty volume'' as considered by Wootters and Fields)
depends on $\tau$, and so do the ``best strategies''. After having
arrived at (two variants of) a general variational principle
determining what is a best strategy, we solve the problem of
finding these strategies in very detail in two dimensions ($d=2$),
and discuss some features of the problem in higher dimensions.
\bigskip

In contrast to the scenario considered by Wootters and Fields, our
experimentalist learns from previously obtained data and uses them
to re-adjust the observable measured in the next step, eventually
approaching the best strategy for the unknown state $\tau$. We
show that, in dimensions larger than $2$, the best strategy is
sometimes {\it not} provided by a family of mutually unbiased
observables.
\bigskip

We conclude the paper by giving some remarks on the infinite dimensional case.
\bigskip

\section{Derivation of the probability density}
\setcounter{equation}{0}
\label{Deriv}

Let us begin introducing some notation. The spectral decomposition of any observable
(hermitean linear operator) reads
\be
A = \sum_{a\,\in\,{\rm Sp}(A)} a\, P_{a}\,,
\label{spec}
\ee
where ${\rm Sp}(A)$ denotes the spectrum (set of eigenvalues) of $A$,
and $P_{a}$ are the (unique) hermitean projections onto the respective
eigenspaces (the spectral projections) satisfying $P_{a}P_{b}=0$ for $a\neq b$, and
summing up to the identity operator: $\sum_{a\,\in\,{\rm Sp}(A)} P_{a} = {\mathbf{1}}$.
In a given state (density matrix) $\rho$, the probability to obtain the outcome $a\in{\rm Sp}(A)$
in a measurement of $A$ is given by
\be
w_a(\rho,A) = \langle P_{a} \rangle_\rho \equiv {\rm Tr}(\rho P_{a}),
\label{prob}
\ee
the symbol $\langle\,\,\rangle_\rho$ denoting the expectation value
in the state $\rho$.
\bigskip

Now suppose we are given $n$ copies of a quantum system,
prepared to be in the same -- unknown -- quantum state $\tau$,
and we are allowed to perform a sequence of measurements of $n$ observables $(A_1,A_2,\dots A_n)$,
each on one copy of the system.
This setting guarantees that the outcomes, collectively denoted as
\be
\Lambda\equiv(a_1,a_2,\dots a_n),
\label{data}
\ee
are statistically independent of each other. Given these data -- what can we say about the state?
This is a case for an application of Bayes' Theorem of elementary
probability theory:
Given a domain $D$ in the space ${\cal S}$ of states, we ask for a probability
that the measurement outcomes $\Lambda$ arise from a state contained in $D$.
In other words, we ask for a probability distribution describing the likelihood
of $\rho$ to be responsible for the experimental data.
This requires the assumption of an {\it \`{a} priori} likelihood, i.e.
a probability measure on ${\cal S}$. A natural candidate is the
measure ${\cal D}\!\rho$ induced by the Hilbert-Schmidt
geometry -- see (\ref{HS}) below --,
but in order to be open for different choices, we include an additional density $\mu(\rho)$.
We will see that things do not depend heavily on this quantity.
Whatever choice is made, Bayes' Theorem tells us that
the desired probability distribution on the space of states is given by
\be
p_\Lambda(\rho)\,=\,C\,\mu(\rho)\,\prod_{j=1}^n\,w_{a_j}(\rho,A_j)
\,\equiv\,C\,\mu(\rho)\,\exp\left(\,\sum_{j=1}^n\,\ln w_{a_j}(\rho,A_j)\right),
\label{pLambda}
\ee
where the constant $C$ is chosen such that
\be
\int_{\cal S} {\cal D}\!\rho\,\,p_\Lambda(\rho) \,=\, 1.
\ee
This is the starting point for our analysis.
\bigskip

The probability density (\ref{pLambda}) is defined for any experimental record
$\Lambda$ consisting of all measurement outcomes (\ref{data}). For small $n$, the statistical
fluctuations in the data lead to a strong dependence of $p_\Lambda(\rho)$ on
$\Lambda$. When the number of measurements is increased in an appropriate way,
the fluctuations get suppressed to any desired degree.
The statistical error in the exponent of (\ref{pLambda}) will, roughly estimated,
be of the order $n^{-1/2}$ times the order of the exponent itself.
\bigskip

A particularly simple setup in which the statistical fluctuations may be suppressed in a
controlled way this is to choose a {\it smaller} set of mutually different observables $(B_1,B_2,\dots B_m)$,
$m\ll n$, and repeat each of them sufficiently often.
In other words, the sequence $(A_1,A_2\dots A_n)$ is chosen to be of the form
\be
(B_1,\dots B_1,B_2\dots B_2,\dots B_m\dots B_m).
\label{B}
\ee
If $B_\beta$ is measured $n_\beta$ times ($\sum_{\beta=1}^m n_\beta=n$), the number of
measurements may be scaled up uniformly by simply replacing $n_\beta\rightarrow k\,n_\beta$
for sufficiently large $k$, while $m$ is kept constant.
\bigskip

Before coming to the main part of our derivation, let us describe
the underlying idea. We assume that sufficiently many different
observables have been chosen (details to be specified below), and
for the moment we ignore $\mu(\rho)$ from (\ref{pLambda}). Once it is guaranteed
that the statistical fluctuations are small, most experimental
data (\ref{data}) will render (\ref{pLambda}) very close to a
family of ``typical'' probability distributions. For large $n$, a
typical $p_\Lambda(\rho)$ may well be approximated by a Gaussian,
peaked around some density matrix $\rho_\Lambda$. The latter
represents the ``best guess'' for the unknown state, i.e. for
$\tau$. Some general properties of $p_\Lambda(\rho)$ may be
inferred from the fact that the exponent in (\ref{pLambda}) is a
sum of $n$ statistically independent quantities: As $n$ increases,
the typical error made by estimating the unknown state to be
$\rho_\Lambda$ scales as $n^{-1/2}$. However, the quadratic form
defining the ``shape'' of the Gaussian only depends -- to leading
order -- on $\tau$ and on the sequence of observables chosen, i.e.
it is approximately the same for all data that may reasonably
occur. Hence, the distributions $p_\Lambda(\rho)$ may be viewed as
translated versions of each other. In order to have a manageable
quantity at hand, we pick out the ``average'' distribution
$p(\rho)$, defined by replacing the exponent in (\ref{pLambda}) by
its expectation value (with respect to $\tau$). As we shall work
out below in detail, it is peaked around $\tau$. An
experimentalist having performed $n$ measurements and having
inserted the data (\ref{data}) into (\ref{pLambda}) will thus
obtain a result very close to
$p_\Lambda(\rho)=p(\rho-\rho_\Lambda+\tau)$, with
$\rho_\Lambda=\tau+O(n^{-1/2})$ being the best guess for the
unknown state. The important point is now that the measure of
knowledge (or uncertainty) depends only on the ``shape'' of the
Gaussian (i.e. on the quadratic form in the exponent), but not on
the location $\rho_\Lambda$ of its center. (The typical
$\rho_\Lambda$ occurring in different runs of the experiment are
distributed according to $p(\rho)$). It is in
this sense that $p(\rho)$ is ``typical'' and is asymptotically
approached by $p_\Lambda(\rho)$ as $n\rightarrow\infty$.
\bigskip

We begin our derivation by considering the exponent
in (\ref{pLambda}) as a function of the data $a_1,\dots a_n$ from (\ref{data}).
The probability distribution relevant for any $a_j$ is $w_a(\tau,A_j)$.
Hence, we define
\be
p(\rho)\,=\,C'\,\mu(\rho)\,\exp\left(\,\sum_{j=1}^n\,R_j(\rho)\right),
\label{pC}
\ee
where
\be
R_j(\rho)\,=\,\sum_{a\,\in\,{\rm Sp}(A_j)} w_a(\tau,A_j)\,\ln w_a(\rho,A_j),
\ee
and $C'$ is a normalization constant close to $C$.
Next, we define quantities $\varepsilon_{ja}(\rho)$ by
\be
w_a(\rho,A_j) = \varepsilon_{ja}(\rho) + w_a(\tau,A_j)
\label{epsilon}
\ee
and write
\be
R_j(\rho)\,=\,H_j\,+\,S_j(\rho),
\label{rel}
\ee
where
\be
H_j\,=\,\sum_{a\,\in\,{\rm Sp}(A_j)} w_a(\tau,A_j)\,\ln w_a(\tau,A_j)
\label{Hj}
\ee
is the negative of the Shannon information measure of the probability
distribution $a\mapsto w_a(\tau,A_j)$ -- it may be absorbed into the constant $C'$ --, and
\be
S_j(\rho)\,=\,\sum_{a\,\in\,{\rm Sp}(A_j)} w_a(\tau,A_j)\,\ln\Big(1+\frac{\varepsilon_{ja}(\rho)}{w_a(\tau,A_j)}\Big)
\ee
is -- according to (\ref{rel}) -- the negative of the relative entropy $S_j(\tau|\rho)$ of
$a\mapsto w_a(\tau,A_j)$ to $a\mapsto w_a(\rho,A_j)$.
Assuming $\varepsilon_{ja}(\rho)$ to be small ($\rho$ being close to $\tau$), we can expand
\be
S_j(\rho)\,=\,\sum_{a\,\in\,{\rm Sp}(A_j)} \left(\varepsilon_{ja}(\rho)
-\,\frac{1}{2}\,\,\frac{\varepsilon^2_{ja}(\rho)}{w_a(\tau,A_j)}
+ O\left(\frac{\varepsilon^3_{ja}(\rho)}{w_a^{\,2}(\tau,A_j)}\right)\right).
\label{S}
\ee
The first term vanishes on account of $\sum_{a\,\in\,{\rm Sp}(A_j)} w_a(\rho,A_j)=$
$\sum_{a\,\in\,{\rm Sp}(A_j)} w_a(\tau,A_j)=1$ and the definition (\ref{epsilon}).
The last term is somewhat delicate. It may be neglected if its denominator is non-zero and
the number of measurements  is sufficiently large.  Hence, we would like to have
$w_a(\tau,A_j)\neq 0\,\,\forall\,a\in{\rm Sp}(A_j)$ and $\forall\,\,j=1,\dots n$.
The simplest way to achieve this is to require
\be
w_a(\tau,A)\neq 0\quad\forall\,\,a\in{\rm Sp}(A)
\label{wsjnmhsc}
\ee
for {\it any} observable $A$. With (\ref{prob}), this is equivalent to
${\rm Tr}(\tau P)\neq 0$ for any (non-zero) hermitean projection $P$,
which just states that $\tau$ is invertible, i.e. all eigenvalues of $\tau$
being non-zero. In finite dimensions, this is not a very drastic condition
on the unknown state: It just states that $\tau$ lies
in the {\it interior} of the set ${\cal S}$ of states. From now on, we shall assume this to be the case.
Thus, omitting the last term in (\ref{S}) may be compensated by a correction factor
of the order
\be
1+O\left(\frac{\varepsilon_{ja}(\rho)}{w_a(\tau,A_j)}\right)\,\approx\,
1+O\left(\parallel\!\rho-\tau\!\parallel\,\parallel\!\tau^{-1}\!\parallel\right)
\label{appr}
\ee
or even closer to $1$,
$\parallel\,\parallel$ denoting the operator norm ($\,\parallel\!\!A \!\!\parallel\,= \max_{a\in{\rm Sp}(A)}|a|\,$).
It may therefore be neglected if $\rho$ is sufficiently close to $\tau$.
We will show below that this will be the case in the region of interest.
\bigskip

Upon omitting the last term in (\ref{S}) and re-inserting
$\varepsilon_{ja}(\rho)$ from (\ref{epsilon}), we arrive at the
result that -- for invertible $\tau$ and after sufficiently many
measurements -- the desired probability density is given by
\be
p(\rho)\,=\,K\,\mu(\rho)\,\exp\left(-\,\frac{1}{2}\,\sum_{j=1}^n\,Q(\rho,A_j)\right),
\label{pK}
\ee
where
\be
Q(\rho,A)\,=\,\sum_{a\,\in\,{\rm Sp}(A)}\,\frac{\Big(w_a(\rho,A) - w_a(\tau,A)\Big)^2}{w_a(\tau,A)}.
\label{Q0}
\ee
With $P_a$ denoting the spectral projections of $A$,
this may also be written as
\be
Q(\rho,A)\,=\,\sum_{a\,\in\,{\rm Sp}(A)}\,\frac{{\rm Tr}^{\,2}\Big((\rho-\tau) P_{a}\Big)}{{\rm Tr}(\tau P_{a})}
\,\equiv\,\sum_{a\,\in\,{\rm Sp}(A)}\,
\frac{\Big(\langle P_{a} \rangle_\rho-\langle P_{a} \rangle_\tau\Big)^2}{\langle P_{a} \rangle_\tau},
\label{Q}
\ee
a quantity which plays a key role in what follows.
The constant $K$ in (\ref{pK}), collecting $C'$ and the $\rho$-independent contribution
(\ref{Hj}), is chosen such that
\be
\int_{\cal S} {\cal D}\!\rho\,\,p(\rho) \,=\, 1.
\label{norm}
\ee
The sum over the $Q$'s in the exponent of (\ref{pK}) defines a quadratic form on ${\cal S}$.
It may be written as
\be
M(\rho) \,\equiv\, \sum_{j=1}^n Q(\rho,A_j)\,=\,(\rho-\tau|{\cal
M}|\rho-\tau), \label{M} \ee where ${\cal M}$ is a linear operator
acting on ${\cal B}_0$, the (real) vector space of hermitean
linear operators with zero trace. Here $(...|...)$ denotes the
Hilbert-Schmidt inner product
\be
(\xi|\eta) \,=\, 2\,{\rm Tr}(\xi^\dagger\eta)
\label{HS}
\ee
for
arbitrary linear operators $\xi$ and $\eta$, which induces a
(real) inner product on ${\cal B}_0$. 
(The factor $2$ is just for
convenience. It ensures that for $d=2$ the matrices $\sigma_{r}/2$
form an orthonormal basis of ${\cal B}_0$). With respect to (\ref{HS}), the
operator ${\cal M}$ is symmetric. We assume that there are enough
independent observables among the $A_j$ so as to make ${\cal M}$
invertible. (In fact, the overall set of all spectral projections
$\{P_{ja}\}$ must span the complete $d^2$-dimensional space of
hermitean linear operators). Hence, $M(\rho)$ is a non-degenerate
quadratic form, the exponential part in (\ref{pK}) being a
distribution of Gaussian type peaked around $\tau$. When the
number of measurements is increased, the peak becomes arbitrarily
sharp, eventually coming to lie well inside the domain in which
(\ref{appr}) may be replaced by $1$. To see this in more detail,
we consider a ``typical'' $\rho$, whose ``distance'' from $\tau$
corresponds to the RMS (root mean square) deviation of the
Gaussian
\be
\parallel\!\rho_{\rm typical}-\tau\!\parallel^2 \,\leq\, {\rm Tr}\left((\rho_{\rm typical}-\tau)^2\right)
\,\approx\, {\rm Tr}({\cal M}^{-1})\,.
\label{app2}
\ee
(For the second step, cf. (\ref{D2}) below).
For large $n$, ${\rm Tr}({\cal M}^{-1})$ becomes proportional to $n^{-1}$.
Hence, $n$ may be chosen large enough so as to make
(\ref{appr}) arbitrarily close to $1$ for any ``typical'' $\rho$.
With increasing $n$, the approximation becomes arbitrarily accurate.
\bigskip

In case of choosing $m$ observables $B_1,\dots B_m$ according to
the scheme (\ref{B}), and performing $n_\beta$ measurements of
$B_\beta$, (\ref{Q}) and (\ref{M}) combine into
\be
M(\rho)\,=\,\sum_{\beta=1}^m\,n_\beta\,\,Q(\rho,B_\beta)\,=\,(\rho-\tau|{\cal
M}|\rho-\tau). \label{MM} \ee In order to render ${\cal M}$
invertible, we must have $m\geq d+1$. The lower bound $m=d+1$ may
only be attained if the overall set of all spectral projections
$\{P_{\beta b}\}$ spans the (real) vector space of hermitean
linear operators, which implies that each $B_\beta$ has only
non-degenerate eigenvalues. This formula will turn out
particularly useful later on.
\bigskip

Now we have to say some words about the {\it \`{a} priori} probability distribution
$\mu$ contained in (\ref{pK}). We mainly focus on situations where nothing -- or very few --
is known about $\tau$ {\it before} the measurements are carried out.
One would then choose $\mu$ to be spread over the whole of ${\cal S}$.
Consequently, the functional dependence of $\mu(\rho)$ is dominated by the peak of
the Gaussian. In particular, if $\mu$ is continuous at $\rho=\tau$,
$\mu(\rho)$ may effectively be replaced by $\mu(\tau)$ for large $n$. Hence, it is justified to ignore this factor,
and we will set $\mu(\rho)\equiv1$ for the rest of this paper.
\bigskip

Finally, the region of integration in (\ref{norm}) may effectively be replaced by
the set of hermitean linear operators with trace unity, which is isomorphic to $\realnumb^{d^2-1}$.
Thus we end up with the standard normalized Gaussian
\be
p(\rho) \,\,=\,\, \sqrt{\frac{{\rm det}{\cal
M}}{(2\pi)^{d^2-1}}}\,\,
\exp\left(-\,\frac{1}{2}\,(\rho-\tau|{\cal M}|\rho-\tau)\right),
\label{pG} \ee where ${\cal M}$ is the linear operator ${\cal M}$
as defined in (\ref{M}) or in the more convenient form (\ref{MM}).
This operator -- depending only on $\tau$ and on the sequence of
observables -- is thus the key object allowing us to quantify the
gain of knowledge in terms of a single numerical measure. We recall that, when the
experimentalist inserts the measurement outcome data (\ref{data})
into (\ref{pLambda}), he will obtain a probability distribution
very close to a translated version of $p(\rho)$, i.e.
\be
p_\Lambda(\rho) \,\,=\,\, \sqrt{\frac{{\rm det}{\cal
M}}{(2\pi)^{d^2-1}}}\,\,
\exp\left(-\,\frac{1}{2}\,(\rho-\rho_\Lambda|{\cal
M}|\rho-\rho_\Lambda)\right),
\ee
where $\rho_\Lambda$ differs from
$\tau$ by $O(n^{-1/2})$.
\bigskip

\section{Measures of knowledge and best strategies in general}
\setcounter{equation}{0}
\label{knowl}

The distribution (\ref{pG}) is determined by the quadratic form
(\ref{M}) or (\ref{MM}), i.e. by the linear operator ${\cal M}$ on the (real)
vector space of hermitean linear operators with trace $0$. As
described above, ${\cal M}$ contains all the information necessary
to work out the experimentalist's knowledge (or uncertainty) about
the unknown state, once he knows the data. The only freedom that
is left for him is to choose the sequence of observables $A_j$.
However, prior to searching a strategy (i.e. a choice of
observables) that maximizes this knowledge, we first have to
specify how the ``knowledge about the unknown state'' -- or,
conversely, the ``uncertainty about the unknown state'' -- is
quantified. The answer depends on which feature of the unknown
state is required. We consider two possible approaches:
\bigskip

\noindent {\bf a.) Volume in ${\cal S}$}:\\
The peak of the Gaussian (\ref{pG}) occupies a ``volume'' in the set ${\cal S}$ of states of the order
\be
{\cal V}\,=\,({\rm det}{\cal M})^{-1/2}\,, \label{V}
\ee
which may be considered as a measure of uncertainty about $\tau$.
This is not identical with, but plays a similar role as
Wootters and Fields' ``uncertainty volume'' \cite{WoottersFields1989}, {\it before} the average over the possible unknown states is performed.
It corresponds to the information theoretic notion of knowledge since
it is related monotonously to the negative of the Shannon
information measure
\be
H \,=\, \int_{\cal S} {\cal D}\!\rho\,\,p(\rho)\,\ln p(\rho)\,=\, -\,\frac{d^2-1}{2} +\,\frac{1}{2}\,\ln\left(\frac{{\rm det}{\cal M}}{(2\pi)^{d^2-1}}\right).
\ee
A best strategy based on this measure (a best ``volume oriented strategy'') is one for which
${\rm det}{\cal M}$ is {\it maximal} for given $n$.
\bigskip

\noindent {\bf b.) Distance from $\tau$}:\\
The RMS (root mean square) deviation of the distribution (\ref{pG}) is given by
\be
D^2\,=\,(\Delta \rho)^2 \,\equiv\, \int_{\cal S} {\cal D}\!\rho\,\,p(\rho)\,\,{\rm Tr}\left((\rho-\tau)^2\right)
\,=\,{\rm Tr}({\cal M}^{-1})\,.
\label{D2}
\ee
It represents the uncertainty about the unknown state as measured in terms of the
mean ``distance squared'' $\,{\rm Tr}\left((\rho-\tau)^2\right)$ in the space ${\cal S}$ of states
and defines a ``length'' scale $D$.
A best strategy based on this measure (a best ``distance oriented strategy'') is one for which
${\rm Tr}\left({\cal M}^{-1}\right)$ is {\it minimal} for given $n$.
\bigskip

It is easy to see that any best strategy based on maximizing ${\rm
det}{\cal M}$ or minimizing ${\rm Tr}\left({\cal M}^{-1}\right)$
necessarily has to use observables $A_j$ with non-degenerate
eigenvalues only. At the level of our
formalism, this feature may be traced back to the properties of
the quadratic form $M(\rho)$, as given by (\ref{M}) or (\ref{MM}),
and its constituents $Q(\rho,A)$ as defined in (\ref{Q0}) and
(\ref{Q}): We first note that $M(\rho)$ is a {\it sum}, each term
stemming from a particular measurement. $Q(\rho,A)$ may thus be
considered as a measure of how our knowledge increases (on the
average) by a measurement of $A$. $M(\rho)$ has the important
property that the contribution of an observable $A$ will be the
larger, the more spectral projections $A$ possesses: Let $A$ be
one of our observables measured, and suppose it possesses a
degenerate eigenvalue $a$. The corresponding eigenspace (the image
of the spectral projection $P_a$) thus has dimension greater than
$1$. Suppose now that the measurement of $A$ is replaced by the
measurement of another observable $A'$, constructed from $A$ by
replacing $aP_a\rightarrow a'P_{a'} +a''P_{a''}$ in the spectral
decomposition of $A$ (where $a'\neq a''$, both numbers being
different from the other eigenvalues of $A$, and $P_{a'}$,
$P_{a''}$ being orthogonal projections dividing the eigenspace
into a direct sum: $P_a = P_{a'}+P_{a''}$). We can consider $A'$
as a ``refinement'' of $A$. Now we compare the two corresponding
quantities $M(\rho)$ and $M'(\rho)$. Explicit computation reveals
\begin{eqnarray}
&&\qquad\quad M'(\rho)-M(\rho)\,=\,Q(\rho,A')-Q(\rho,A) \,=\nonumber\\
&&\frac{\Bigg(
{\rm Tr}(\tau P_{a'})\,
{\rm Tr}\Big((\rho-\tau) P_{a''}\Big) -
{\rm Tr}(\tau P_{a''})\,
{\rm Tr}\Big((\rho-\tau) P_{a'}\Big)
\Bigg)^2}{{\rm Tr}(\tau P_{a'})\,{\rm Tr}(\tau P_{a''})\,{\rm Tr}\Big(\tau (P_{a'}+P_{a''})\Big)}\,,
\end{eqnarray}
which represents a semi-positive quadratic form by its own.
Consequently, we have ${\rm det}{\cal M}'\geq{\rm det}{\cal M}$ and
${\rm Tr}({\cal M}'^{-1})\leq{\rm Tr}({\cal M}^{-1})$, while
the total number $n$ of measurements has not been changed. The same procedure may be
repeated until all degenerate eigenvalues of all observables $A_j$ have disappeared.
(The same behaviour is expected for any other reasonable measure of knowledge).
\bigskip

By construction, the best strategies depend on the unknown state
$\tau$. Hence, one may object that when $\tau$ is {\it unknown},
the experimentalist does not know how to choose his observables.
On the other hand, when inserting the outcomes of a relatively
small number of measurements of arbitrary observables into
(\ref{pLambda}), one obtains a first rough estimate of $\tau$.
Next, one chooses observables according to a best strategy {\it as
if} the estimate in fact coincides with the unknown state. After
some runs of this type (or even after each measurement) one
determines a better estimate for $\tau$ and re-adjusts the
observables. This procedure is iterated and will, for increasing
$n$, approach the effectiveness  of a best strategy.
\bigskip

\section{The two-dimensional case}
\setcounter{equation}{0}
\label{two-dim}

Let us now study the case $d=2$ in some detail.
Since the states and observables on a two-dimensional Hilbert space admit
a simple geometric representation, we can be more explicit than in the case of general $d$.
The set of all density matrices may be parametrized as
\be
\rho(\vec{a}) = \frac{1}{2}\left({\mathbf{1}} + \vec{a}\,\vec{\sigma}\right)
\qquad{\rm with}\qquad |\vec{a}|\leq 1,
\label{rho}
\ee
where $\vec{\sigma}$ represents
three observables obeying the Pauli spin matrix algebra, and $\vec{a}\in\realnumb^3$.
Pure states are characterized by $|\vec{a}|=1$, the tracial state is given by $\vec{a}=0$.
The space ${\cal S}$ of states is thus represented by the unit ball in $\realnumb^3$.
The natural measure ${\cal D}\!\rho$ on ${\cal S}$ is the Euclidean volume element $d^3a$.
\bigskip

Now let us look at observables. Any hermitean linear operator may be written as $a {\mathbf{1}} + \vec{c}\,\vec{\sigma}$
with $a\in\realnumb$ and $\vec{c}\in\realnumb^3$. Leaving aside multiples of the identity and
irrelevant multiplicative factors, we confine ourselves to measuring observables of the type
\be
B(\vec{c}) = \vec{c}\,\vec{\sigma}\qquad{\rm with}\qquad |\vec{c}|=1.
\label{obs2_2}
\ee
The spectrum of any such operator is $\{-1,1\}$. The spectral projection corresponding to the eigenvalue $b\in\{-1,1\}$
of $B(\vec{c})$ takes the convenient form $P_{b} = \frac{1}{2}({\mathbf{1}} + b\, \vec{c}\,\vec{\sigma})$,
and the measurement outcome probabilities for this observable in the state $\rho(\vec{a})$ read
\be
w_b(\rho(\vec{a}),B(\vec{c})) = \frac{1}{2}\left(1 + b\,\vec{a}\,\vec{c}\,\right).
\label{w_2}
\ee
We now specify our sequence of observables according to the scheme (\ref{B}): We choose $m$ unit vectors
$\vec{c}_\beta$ ($\beta=1,\dots m$) and perform $n_\beta$ measurements of each
$B_\beta\equiv B(\vec{c}_\beta)$. The total number of measurements is therefore $n=\sum_{\beta=1}^m n_\beta$.
The unknown state shall be represented the parameter value $\vec{u}$, i.e.
\be
\tau\equiv\rho(\vec{u}),
\ee
and $p(\rho)$ is written as $p(\vec{a})$. Using (\ref{MM}),
a short computation reveals that the probability distribution (\ref{pG}) is given by
\be
p(\vec{a}) \,=\, \sqrt{\frac{{\rm det}{\cal M}}{(2\pi)^3}}\,\,
\exp\left(-\,\frac{1}{2}\,(\vec{a}-\vec{u})^T {\cal M}\,(\vec{a}-\vec{u})\,\right),
\label{pp2}
\ee
where ${\cal M}$ is the $3\times 3$ matrix with components
\be
{\cal M}_{rs} \,=\,\sum_{\beta=1}^m\,\,\frac{n_\beta}{1-(\vec{u}\,\vec{c}_\beta)^2}\,\,\,c_{\beta r}\,c_{\beta s},
\label{MM2comp}
\ee
$c_{\beta r}$ being the components of the vector $\vec{c}_\beta$, with $r$ and $s$ ranging from $1$ to $3$.
Starting with this expression, we will now tackle the problem of finding all best strategies for
the determination of the unknown state $\rho(\vec{u})$.
\bigskip

\section{Best strategies for $d=2$}
\setcounter{equation}{0}

The form of (\ref{MM2comp}) shows that we must have $m\geq 3$, and the sequence of vectors
$\vec{c}_\beta$ must contain three linearly independent elements (otherwise
${\cal M}$ would not be invertible).
In other words, the $m\times 3$ matrix defined by the components $c_{\beta r}$ must have
rank 3. According to the two measures of knowledge
as discussed in section \ref{knowl}, we consider the two cases of
maximizing ${\rm det}{\cal M}$ and minimizing ${\rm Tr}({\cal M}^{-1})$.
\bigskip

\noindent {\bf a.) Maximizing ${\rm det}{\cal M}$}:\\
We first consider the volume oriented approach, i.e. the
case when the ``volume'' ${\cal V}$ in ${\cal S}$ occupied by the peak of the Gaussian
(\ref{pp2}) serves as a measure for the uncertainty about the unknown state.
Let us fix $n$ (and, for the moment, $m$) and ask for which configurations $(n_\beta,\vec{c}_\beta)$
the determinant of ${\cal M}$ is maximal
under the subsidiary conditions $\sum_{\beta=1}^m n_\beta = n$ and $|\vec{c}_\beta|=1\,\,\forall\beta$.
Introducing Lagrange multipliers $c$, $C_\beta$, the corresponding unconstrained problem is to maximize
\be
{\cal F} \,=\, \ln {\rm det}{\cal M} - c \sum_{\beta=1}^m n_\beta - \frac{1}{2}\,\sum_{\beta=1}^m C_\beta\,\vec{c}_\beta^{\,2}
\ee
with respect to the variables $(n_\beta,\vec{c}_\beta,c,C_\beta)$.
The logarithm is used just for convenience: This form allows us to apply the general formula
$\partial(\ln {\rm det}{\cal M}) = {\rm Tr}({\cal M}^{-1}\partial{\cal M})$, where $\partial$ stands
for any derivative $\partial/\partial c_{\beta r}$ or $\partial/\partial n_\beta$.
Now we choose the coordinates in $\realnumb^3$ such that
${\cal M}$ is diagonal in the maximizing configuration.
This choice is possible because ${\cal M}$ is a hermitean matrix,
and it causes all non-diagonal elements to drop out of the problem.
Differentiation with respect to $n_\beta$ and $c_{\beta r}$ leads to the set of equations
\begin{eqnarray}
\frac{1}{1-(\vec{u}\,\vec{c}_\beta)^2}\,\,
\sum_{s=1}^3 \,\frac{c_{\beta s}^{\,\,2}}{{\cal M}_{ss}}&=& c\label{e1}\\
\frac{1}{1-(\vec{u}\,\vec{c}_\beta)^2}\,\,
\frac{2\,n_\beta\,c_{\beta r}}{{\cal M}_{rr}} \,+\,
\frac{2\,n_\beta (\vec{u}\,\vec{c}_\beta)\,u_r}{\Big(1-(\vec{u}\,\vec{c}_\beta)^2\Big)^2}\,\,
\sum_{s=1}^3\,\frac{c_{\beta s}^{\,\,2}}{{\cal M}_{ss}}
&=& C_\beta\, c_{\beta r}\,,
\label{e2}
\end{eqnarray}
whose combination yields
\be
\frac{2\,n_\beta}{1-(\vec{u}\,\vec{c}_\beta)^2}
\left(\,\frac{c_{\beta r}}{{\cal M}_{rr}} \,+\, (\vec{u}\,\vec{c}_\beta)\,u_r\,c
\right) \,=\, C_\beta\, c_{\beta r}\,.
\label{e3}
\ee
Multiplying this equation by $c_{\beta r}$, summing over $r$ and using (\ref{e1})
and $|\vec{c}_\beta|=1\,\,\forall\beta$ gives
\be
C_\beta \,=\, \frac{2\,n_\beta \,c}{1-(\vec{u}\,\vec{c}_\beta)^2}\,.
\label{eC}
\ee
Multiplying (\ref{e1}) by $n_\beta$, summing over $\beta$ and using (\ref{MM2comp}) and
$\sum_{\beta=1}^m n_\beta = n$ leads to
\be
c\,=\,\frac{3}{n}\,.
\label{ec}
\ee
Upon inserting these last two expressions into (\ref{e3}), we find
\be
c_{\beta r} \left(\,\frac{1}{{\cal M}_{rr}}\,-\,\frac{3}{n}\,\right)\,\,+\,\,\frac{3}{n}\,\,(\vec{u}\,\vec{c}_\beta)\,\,u_r
\,=\,0\,.
\label{e4}
\ee
Since the $m\times 3$ matrix defined by $c_{\beta r}$ has rank 3 -- as argued
at the beginning of this section --, the term
$({\cal M}_{rr})^{-1}-3\,n^{-1}$ must vanish for at least two values of $r$
(otherwise one could divide by these terms for two or three values of $r$
and conclude that $c_{\beta r}$ has rank less than 3). We may choose
the coordinates of $\realnumb^3$ such that these values are $r=1$ and $2$.
Hence, ${\cal M}_{11}={\cal M}_{22}=\frac{1}{3}n$, which implies
$u_1=u_2=0$, $\vec{u}\,\vec{c}_\beta=u_3\,c_{\beta 3}$ and $u_3^2=\vec{u}^2$.
Equation (\ref{e4}) thus shrinks to the statement that
${\cal M}_{33}=\frac{1}{3}n(1-\vec{u}^2)^{-1}$, and
the remaining equation (\ref{e1}) is automatically satisfied.
In this way we arrive at the following
\bigskip

\noindent {\bf Lemma 1}:\\
For given $\vec{u}$ and $n$, the configuration $(m,n_\beta,\vec{c}_\beta)$
 maximizes ${\rm det}{\cal M}$
under the subsidiary conditions $\sum_{\beta=1}^m n_\beta = n$ and
$|\vec{c}_\beta|=1\,\,\forall\beta$ if and only if

{\it (i)} $\vec{u}$ is an eigenvector of ${\cal M}$ associated with the
          eigenvalue $\frac{1}{3}n(1-\vec{u}^2)^{-1}$, and

{\it (ii)} the two other eigenvalues of ${\cal M}$ are both equal to $\frac{1}{3}n$.

\noindent The second statement implies that
${\cal M}$ acts proportional to the identity in the subspace orthogonal to $\vec{u}$.
The value of ${\rm det}{\cal M}$ in the maximizing configuration is given by
\be
({\rm det}{\cal M})_{\rm max}
\,=\,\left(\frac{n}{3}\right)^3\frac{1}{1-\vec{u}^2}\,,
\label{detmax} \ee or, expressed in terms of the ``volume'' ${\cal
V} = ({\rm det}{\cal M})^{-1/2}$ occupied by the peak of the
Gaussian,
\be
{\cal V}_{\rm min} \,=\,\left(\frac{3}{n}\right)^{3/2}\sqrt{1-\vec{u}^2}\,.
\label{Volmin}
\ee
\bigskip

Any configuration satisfying {\it (i)} and {\it (ii)} represents a ``best strategy'',
and all these strategies work equally well, because (\ref{detmax}) depends only on $n$
and $\vec{u}$, but not on any details of the configuration $(m,n_\beta,\vec{c}_\beta)$.
The simplest strategy is to choose $m=3$ and let
$\{\vec{c}_1,\vec{c}_2,\vec{c}_3\}$ be an orthonormal basis of $\realnumb^3$
such that one of these vectors ($\vec{c}_3$, say) is parallel to $\vec{u}$.
In this strategy, we must have $n_1=n_2=n_3=\frac{1}{3}n$, i.e. all three observables
$B_\beta\equiv B(\vec{c}_\beta)$ are measured equally often.
\bigskip

\noindent {\bf b.) Minimizing ${\rm Tr}({\cal M}^{-1})$}:\\
The distance oriented approach, i.e. the case when the mean ``distance squared'' from the center of the Gaussian
(\ref{pp2}) serves as a measure for the uncertainty about the unknown state,
is treated similarly. Formally, the problem consists of minimizing
\be
{\cal F} \,=\, {\rm Tr}({\cal M}^{-1}) + c \sum_{\beta=1}^m n_\beta + \frac{1}{2}\,\sum_{\beta=1}^m C_\beta\,\vec{c}_\beta^{\,2}\,,
\ee
where $c$ and $C_\beta$ are Lagrange multipliers.
We again choose the coordinates in $\realnumb^3$ such that ${\cal M}$ is diagonal in the minimizing configuration
and use the general formula
$\partial({\rm Tr}({\cal M})^{-1}) = -\,{\rm Tr}({\cal M}^{-1}(\partial{\cal M}){\cal M}^{-1})$,
where $\partial$ stands
for $\partial/\partial c_{\beta r}$ and $\partial/\partial n_\beta$.
Differentiation yields a set of equations that look like (\ref{e1})--(\ref{e2}),
except that the diagonal elements ${\cal M}_{rr}$ and ${\cal M}_{ss}$ are replaced by their squares,
and the same applies to the analogue of (\ref{e3}). Equation (\ref{eC}) appears
without change, but the analogue of (\ref{ec}) now takes the form
\be
c\,=\,\frac{1}{n}\,\sum_{s=1}^3\frac{1}{{\cal M}_{ss}}\,,
\label{ecc}
\ee
due to an additional appearance of $({\cal M}_{ss})^{-1}$ in the analogue of (\ref{e1}).
Hence, the analogue of (\ref{e4}) becomes
\be
c_{\beta r} \left(\,\frac{1}{{\cal M}_{rr}^{\,\,2}}\,-\,c\right)\,\,+\,\,c\,(\vec{u}\,\vec{c}_\beta)\,\,u_r\,=\,0
\ee
with $c$ from (\ref{ecc}). Following the same logic as before, the term
$({\cal M}_{rr})^{-2}-c$ must vanish for at least two values of $r$
(which we choose to be $1$ and $2$). This implies $u_1=u_2=0$ and
\be
\frac{1}{{\cal M}_{11}^{\,\,2}}\,=\,\frac{1}{{\cal M}_{22}^{\,\,2}}=c
\qquad
\frac{1}{{\cal M}_{33}^{\,\,2}} \,=\, c\,(1-\vec{u}^2)\,.
\ee
Combining these equations with (\ref{ecc}), we may easily
compute the diagonal elements ${\cal M}_{rr}$, i.e. the eigenvalues
of ${\cal M}$ (to be displayed below).
The remaining equation -- the analogue of (\ref{e1}) -- is then automatically satisfied.
Our result thus reads:
\bigskip

\noindent {\bf Lemma 2}:\\
For given $\vec{u}$ and $n$, the configuration $(m,n_\beta,\vec{c}_\beta)$
 minimizes ${\rm Tr}({\cal M}^{-1})$
under the subsidiary conditions $\sum_{\beta=1}^m n_\beta = n$ and
$|\vec{c}_\beta|=1\,\,\forall\beta$ if and only if

{\it (i)} $\vec{u}$ is an eigenvector of ${\cal M}$ associated with the eigenvalue
\be
\frac{n}{(2+\sqrt{1-\vec{u}^2})\sqrt{1-\vec{u}^2}}\,,
\ee
\indent and

{\it (ii)} the two other eigenvalues of ${\cal M}$ are both equal to
\be
\frac{n}{2+\sqrt{1-\vec{u}^2}}\,.
\ee

\noindent
The second statement implies that
${\cal M}$ acts proportional to the identity in the subspace orthogonal to $\vec{u}$.
The value of ${\rm Tr}({\cal M}^{-1})$ in the minimizing configuration is given by
\be
D^2_{\rm min} \,\equiv \,{\rm Tr}({\cal M}^{-1})_{\rm min} \,=\,\frac{1}{n}\,\left(2+\sqrt{1-\vec{u}^2}\,\right)^2\,.
\label{Trmin}
\ee
\bigskip

Any configuration satisfying {\it (i)} and {\it (ii)} represents a ``best strategy'', and all these strategies
work equally well. The simplest one is to choose $m=3$ and let
$\{\vec{c}_1,\vec{c}_2,\vec{c}_3\}$ be an orthonormal basis of $\realnumb^3$
such that one of these vectors ($\vec{c}_3$, say) is parallel to $\vec{u}$.
The numbers of measurements performed of any of the three observables
$B_\beta\equiv B(\vec{c}_\beta)$ must now be chosen as
\be
n_1\,=\,n_2\,=\,\frac{n}{2+\sqrt{1-\vec{u}^2}}
\qquad
n_3\,=\,\frac{n\,\sqrt{1-\vec{u}^2}}{2+\sqrt{1-\vec{u}^2}}\,,
\ee
and they correctly sum up to $n$. The observable aligned with
$\vec{u}$ thus needs less measurements than the others.
\bigskip

\noindent {\bf Comparing a.) and b.)}:\\ The knowledge about the
unknown state after $n$ optimally chosen measurements is given by
the volume (\ref{Volmin}) and the length squared (\ref{Trmin}),
respectively. For small $|\vec{u}|$, these two methods work
roughly equally well. In both cases, the three eigenvalues of
${\cal M}$ are approximately of the same order, the spread of the
Gaussian thus being roughly the same in all directions in ${\cal
S}$. If, however, $|\vec{u}|$ is close to $1$ (i.e. $\tau$ being
almost pure), one eigenvalue of ${\cal M}$ becomes large in both
cases, thus causing the peak to be spread only very little in the
direction of $\vec{u}$. In this situation the ``volume'' oriented
approach is more efficient: In the limit $|\vec{u}|\rightarrow 1$
for fixed $n$ we have ${\cal V}_{\rm min}\rightarrow 0$, whereas
$D^2_{\rm min}\rightarrow 4n^{-1}$.
\bigskip

In both cases, the strategy works as follows:
When inserting the outcomes of a relatively small number of measurements
 of arbitrary observables into (\ref{pLambda}), one obtains a first rough
  estimate of $\tau$, i.e. of $\vec{u}$. Next one chooses an orthonormal basis
$\{\vec{c}_1,\vec{c}_2,\vec{c}_3\}$ of $\realnumb^3$ such that $\vec{c}_3$
 is parallel to the best guess of $\vec{u}$. One then measures the three
 corresponding observables $B(\vec{c}_\beta)$ (the relative number of measurements
 depending on whether ${\cal V}$ or $D^2$ represents the measure of uncertainty).
  After some runs of this type (or even after each measurement) one
determines a better guess of $\vec{u}$ and re-adjust the three vectors accordingly
($\vec{c}_3^{\rm \,\,new}$ being aligned with the new guess of $\vec{u}$,
and $\vec{c}_1^{\rm \,\,new}$ and $\vec{c}_2^{\rm \,\,new}$ being as close to
$\vec{c}_1$ and $\vec{c}_2$ as possible). This procedure is iterated and will,
for increasing $n$, converge to an orthonormal basis representing a ``best strategy''
 as determined above. In other words: for sufficiently large $n$, we expect the bounds
  (\ref{Volmin}) or (\ref{Trmin}), respectively, to be approached arbitrarily well.
\bigskip

\section{Comparison of strategies in higher dimensions}
\setcounter{equation}{0}

In this section we consider the case of higher dimensional Hilbert
spaces. After presenting a generally applicable method to improve
strategies, we show how concrete strategies may be constructed. It
turns out that a strategy based on mutually unbiased
(complementary) observables is not always optimal. Concluding, we
give some remarks about the infinite dimensional case.
\bigskip
\bigskip

\noindent {\bf General formalism}
\medskip

\noindent We now turn to higher dimensions. Let the dimension $d$
of ${\cal H}$ be arbitrary. By ${\cal B}$, we denote the (complex)
vector space of all linear operators on ${\cal H}$, endowed with
the Hilbert-Schmidt inner product (\ref{HS}). The latter makes
${\cal B}$ a $d^2$-dimensional Hilbert space by its own. We will
use a bra-ket-notation for this space, using round brackets, i.e.
$|\xi)(\eta|$ representing the linear operator ${\cal
B}\rightarrow {\cal B}$ sending $\zeta \mapsto |\xi)(\eta|\zeta)$
or, equivalently, $\zeta \mapsto 2\,{\rm
Tr}(\eta^\dagger\zeta)\,\xi$. The determinant and trace of linear
operators ${\cal B}\rightarrow {\cal B}$ will be denoted by the
symbols ${\rm det}_\odot$ and ${\rm Tr}_\odot$, respectively.
Furthermore, we need a component formalism for operators of this
type. If $\{e_I|I=1,\dots d\}$ is an orthonormal basis of ${\cal
H}$, the linear operators (``matrix units'')
\be
e_{IJ} \equiv |e_I\rangle\langle e_J|: \,{\cal H}\rightarrow{\cal H}
\label{matrixunits}
\ee
form a basis of ${\cal B}$, satisfying $(e_{IJ}|e_{KL})=2\,\delta_{IK}\delta_{JL}$.
Along with the expansion of elements $\xi\in{\cal B}$ as
\be
\xi = \sum_{I,J} \xi_{IJ}e_{IJ} \equiv \sum_{I,J} |e_I\rangle
\xi_{IJ} \langle e_J| \qquad{\rm with}\quad\xi_{IJ} = \langle
e_I|\xi|e_J\rangle, \ee any linear operator ${\cal A}:{\cal
B}\rightarrow{\cal B}$ may be written as
\be
{\cal A} \,=\,\frac{1}{2}\,\sum_{I,J,K,L}|e_{IJ}){\cal A}_{IJ,KL} (e_{KL}|
\qquad{\rm with}\quad{\cal A}_{IJ,KL} = \frac{1}{2}\,(e_{IJ}|{\cal A}|e_{KL})\,.
\ee
In terms of
these components, the action of ${\cal A}$ is represented by a
matrix multiplication. When understanding the values of the double
index $IJ$ by a single index $r$, the components ${\cal
A}_{IJ,KL}$ explicitly define a $d^2\!\times\!d^2$ matrix
representation ${\cal A}_{rs}$, in which the determinant and the
trace take their usual form. If ${\cal A}=|\xi)(\eta|$, we have
\be
{\cal A}_{IJ,KL} \,=\, 2\,\xi_{IJ}\,\,\eta_{KL}^{\,\,*}\,.
\ee
The orthogonal projection onto the normalized element $(2d)^{-1/2}\,{\mathbf{1}}$
\be
{\cal P}\,=\,\frac{1}{2d}\,|{\mathbf{1}})({\mathbf{1}}| : \,{\cal B}\rightarrow{\cal B}
\label{calP}
\ee
(${\mathbf{1}}$ denoting the unit operator on ${\cal H}$) has components
${\cal P}_{IJ,KL} = d^{-1}\delta_{IJ}\,\delta_{KL}$.
\bigskip

By ${\cal B}_0$, we denote the subset of ${\cal B}$ consisting of all hermitean
 linear operators with zero trace.
It is a real vector space of dimension $d^2-1$, and the Hilbert-Schmidt inner product
for any pair of its elements is real.
The determinant and trace of linear operators ${\cal B}_0\rightarrow {\cal B}_0$
 are denoted by the symbols
${\rm det}$ and ${\rm Tr}$, respectively.
\bigskip

We now consider a strategy based on the scheme (\ref{B}), i.e.
a collection $B_\beta$ ($\beta=1,\dots m$) of operators, such that
each $B_\beta$ is measured $n_\beta$ times in a copy of the system, and
$\sum_{\beta=1}^m n_\beta = n$.
We assume $n_\beta\gg 1$ for each $\beta$.
(As noted above, sufficiently large $n$ may be achieved
by replacing $n_\beta\rightarrow k\,n_\beta$
for sufficiently large $k$, while keeping $m$ constant.)
The key object describing the quality of the strategy is the symmetric linear operator
${\cal M}: {\cal B}_0\rightarrow {\cal B}_0$ as defined in (\ref{MM})
and appearing in the Gaussian (\ref{pG}).
As may be read off from (\ref{MM}) and (\ref{Q}), any observable $A$ provides a contribution
\be
{\cal Q}(A)\,\,=\,\,\,\frac{1}{2}\,
\sum_{a\in{\rm Sp}(A)}\frac{|P_a)(P_a|}{(\tau|P_a)}\,,
\label{calQ}
\ee
where $P_a$ is the spectral projection of $A$ with respect to the eigenvalue $a$.
However, when written in the above form, any such object is a hermitean linear
operator ${\cal Q}(A):{\cal B}\rightarrow{\cal B}$ that
does not leave ${\cal B}_0$ invariant. Its components are given by
\be
{\cal Q}_{IJ,KL}(A)\,=\,\sum_{a\in{\rm Sp}(A)}\frac{P_{a,IJ}\,P_{a,KL}^{\,\,*}}{(\tau|P_a)},
\label{calQcomp}
\ee
where $P_{a,IJ}$ are the components of $P_a$.
From now on, we assume the orthornormal basis $\{e_I\}$ to consist of eigenvectors of $\tau$.
As a consequence, the matrix $\tau_{IJ}$ is diagonal, and the denominator in (\ref{calQcomp}) is
$(\tau|P_a)=2\,\sum_I \tau_{II} P_{a,II}$.
\bigskip

When summing up (\ref{calQ}) for the observables $B_\beta$, we arrive at a hermitean operator
acting on ${\cal B}$. It will turn out convenient to generalize it to a family of operators
${\cal M}_\odot(\alpha): {\cal B}\rightarrow {\cal B}$, defined as
\be
{\cal M}_\odot(\alpha) \,=\,\sum_{\beta=1}^m\,n_\beta\,{\cal Q}(B_\beta)\,+\,\alpha\,{\cal P},
\label{MB}
\ee
where ${\cal P}$ is given by (\ref{calP}). Since
$({\mathbf{1}}|\xi)\equiv2\,{\rm Tr}(\xi)=0$ for any
traceless $\xi$, we have $(\xi|{\cal M}_\odot(\alpha)|\eta)=(\xi|{\cal M}|\eta)$ for all
$\xi,\eta\in{\cal B}_0$. This establishes the relation between ${\cal M}_\odot(\alpha)$
 and the original object ${\cal M}:{\cal B}_0\rightarrow{\cal B}_0$.
\bigskip

We will now express our two measures of knowledge, (\ref{V}) and (\ref{D2}),
in terms of ${\cal M}_\odot(\alpha)$.
Since ${\cal P}$ is the (one-dimensional) hermitean projection onto the
orthogonal complement of ${\cal B}_0$, (\ref{MB}) tells us that
\be
{\rm det}_\odot{\cal M}_\odot(\alpha) = \alpha\,{\rm det}{\cal M} +{\rm terms\,\,independent\,\,of\,\,}\alpha.
\ee
From this it follows
\be
{\rm det}{\cal M} \,=\,\lim_{\alpha\rightarrow\infty}\frac{{\rm det}_\odot{\cal M}_\odot(\alpha)}{\alpha},
\label{detM}
\ee
and analogously we conclude
\be
{\rm Tr}({\cal M}^{-1})\,=\,\lim_{\alpha\rightarrow\infty}{\rm Tr}_
\odot\left({\cal M}_\odot(\alpha)^{-1}\right).
\label{TrinvM}
\ee
These two quantities is all we need in order to compare strategies.
\bigskip
\bigskip

\noindent {\bf Improving strategies}
\medskip

\noindent We now return to the problem of optimizing measurement strategies.
Given some particular strategy characterized by ${\cal M}_\odot(\alpha)$,
we show how to construct another strategy which is at least as good as the original one.
\bigskip

For any observable $B_\beta$, we consider the family
$B^{\,'}_\beta(\varphi) = e^{i\varphi\tau}B_\beta\,e^{-i\varphi\tau}$.
We may think of unitarily ``rotating'' $B_\beta$ within ${\cal B}$ in such a way that the unknown state
$\tau$ is invariant.
When selecting an arbitrary value of $\varphi$, and replacing the
observables $B_\beta$ by $B^{\,'}_\beta(\varphi)$, we obtain a
strategy that is obviously equivalent to the original one.
Denoting its associated family of ${\cal M}$-operators by ${\cal
M}^{\,'}_{\odot}(\alpha,\varphi)$, we have
\begin{eqnarray}
{\rm det}_\odot{\cal M}^{\,'}_{\odot}(\alpha,\varphi) &
=& {\rm det}_\odot{\cal M}_\odot(\alpha),\label{equ1}\\
{\rm Tr}_\odot\left({\cal M}^{\,'}_{\odot}(\alpha,\varphi)^{-1}\right) &
=& {\rm Tr}_\odot\left({\cal M}_\odot(\alpha)^{-1}\right)
\label{equ2}
\end{eqnarray}
for all $\alpha$. We will now construct a further strategy out of
these equivalent ones: We distribute the $n_\beta$ measurements
originally reserved for $B_\beta$ among members of the family
$B^{\,'}_\beta(\varphi)$. Technically, we introduce a probability
distribution $\varphi\mapsto f(\varphi)$ according to which a
value for $\varphi$ is thrown in order to determine the observable
$B^{\,'}_\beta(\varphi)$ to be measured next. In a first step we
may think of $f$ as a discrete distribution (admitting only
particular values for $\varphi$). However, for sufficiently large
$n_\beta$, this may arbitrarily well be approximated by allowing
$f$ to be a continuous distribution. Hence, the average number of
measurements carried out for observables $B^{\,'}_\beta(\varphi)$
satisfying $\varphi_0\leq \varphi\leq\varphi_0+d\varphi$ 
will be $n_\beta f(\varphi_0) d\varphi$.
\bigskip

It may of course happen that different observables $B_\beta$ effectively
 play the same role in the new strategy.
This will happen if they are already ``rotated'' versions of each other, e.g. if
$B_2 = e^{i\varphi\tau}B_1\,e^{-i\varphi\tau}$ and $f(\varphi)\neq 0$ for some $\varphi$.
In this case, the new strategy is effectively generated by a {\it smaller}
 set of observables
than contained in the original strategy (while the number of different observables
actually measured will in general {\it increase}).
\bigskip

By construction, the operator ${\cal M}^{\,'}_\odot(\alpha)$ for the new
strategy is given by the average
\be
{\cal M}^{\,'}_\odot(\alpha) \,\,=\,\,
\int\!d\varphi\,f(\varphi)\,{\cal M}^{\,'}_\odot(\alpha,\varphi).
\label{Maverage} \ee Since ${\cal M}\mapsto{\cal M}^{-1}$ and
${\cal M}\mapsto{\rm ln}\,{\cal M}$ are operator convex functions,
it follows from the Peierls-Bogoliubov inequality that
\begin{eqnarray}
&&{\rm det}_\odot\,{\cal M}^{\,'}_\odot(\alpha) \,\equiv\, {\rm exp}\left({\rm Tr}_
\odot({\rm ln}\,{\cal M}^{\,'}_\odot(\alpha))\right)
\,\ge\, {\rm det}_\odot\,{\cal M}_\odot(\alpha)\,,\\
&&{\rm Tr}_\odot({\cal M}^{\,'}_\odot(\alpha)^{-1}) \,\le\, {\rm Tr}_
\odot({\cal M}_\odot(\alpha)^{-1}),
\end{eqnarray}
where have taken into account (\ref{equ1}) and (\ref{equ2}).
These inequalities survive the limits (\ref{detM}) and (\ref{TrinvM}), so that we conclude
\begin{eqnarray}
{\rm det}\,{\cal M}^{\,'} &\ge& {\rm det}\,{\cal M}\,,\\
{\rm Tr}({\cal M}^{\,'\,-1}) &\le& {\rm Tr}({\cal M}^{-1})\,.
\end{eqnarray}
With respect to the measures of knowledge in both the volume and the distance oriented
approach, the new strategy is better than (or equally well as) the original one.
\bigskip

Let us now compute the operator ${\cal M}^{\,'}_\odot(\alpha)$ for the new
 strategy more explicitly.
The spectral projection of $B^{\,'}_\beta(\varphi)$ with respect to the eigenvalue $a$
is given by $P^{\,'}_{\beta a}(\varphi) = e^{i\varphi\tau}P_{\beta a}\,e^{-i\varphi\tau}$.
Hence, $(\tau|P^{\,'}_{\beta a}(\varphi)) = (\tau|P_{\beta a})$ for any $\varphi$,
 so that nothing changes
in the denominators in (\ref{calQ}) and (\ref{calQcomp}).
Since the basis vectors $e_I$ are eigenvectors of $\tau$, the components of the
 new spectral projections become
$P^{\,'}_{\beta a,IJ}(\varphi)\equiv\langle e_I|P^{\,'}_{\beta a}(\varphi)|e_J\rangle
=e^{i\varphi(\tau_{II}-\tau_{JJ})}\langle e_I|P_{\beta a}|e_J\rangle
\equiv e^{i\varphi(\tau_{II}-\tau_{JJ})} P_{\beta a,IJ}$.
Thus, when computing the components ${\cal M}^{\,\,'}_{\odot,IJ,KL}(\alpha)$,
 the integral over $\varphi$ is to be taken over
\be
e^{i\varphi(\tau_{II}-\tau_{JJ}-\tau_{KK}+\tau_{LL})}\,.
\ee
In order to give it a simple form, we choose $f$ such that the integral over these
expressions is only non-zero if $I=J$ and $K=L$ or $I=K$ and $J=L$. This gives
\be
\int\!d\varphi\,f(\varphi)\,e^{i\varphi(\tau_{II}-\tau_{JJ}-\tau_{KK}+\tau_{LL})}
\,\,=\,\, \delta_{IJ}\,\delta_{KL} + \delta_{IK}\,\delta_{JL} - \delta_{IJKL}\,,
\label{delta}
\ee
where $\delta_{IJKL}=1$ if all four indices agree, and $0$ otherwise.
Strictly speaking, this is only possible if
the eigenvalues $\tau_{II}$ are sufficiently different from each other.
If this is not the case, one may choose some appropriate hermitean operator $\xi$
commuting with $\tau$ and redefine
$B^{\,'}_\beta(\varphi) = e^{i\varphi\xi}B_\beta\,e^{-i\varphi\xi}$.
Coosing the eigenvalues of $\xi$ to have only rational quotients, there is always
 a finite interval for the
$\varphi$-integration such that (\ref{delta}) is valid
with $f(\varphi)={\rm const}$. Otherwise one would have to use the invariant mean
\be
\int\!d\varphi\,f(\varphi) \dots\, \longrightarrow\,
\lim_{T\rightarrow\infty}\frac{1}{2\,T}\int_{-T}^T\!d\varphi\dots
\ee
With the choice (\ref{delta}), the transition from the old to the new strategy is simply achieved by
\be
{\cal M}^{\,\,'}_{\odot,IJ,KL}(\alpha)\,
=\,(\delta_{IJ}\,\delta_{KL} + \delta_{IK}\,\delta_{JL} - \delta_{IJKL})\,
{\cal M}_{\odot,IJ,KL}(\alpha)\,.
\label{Modot}
\ee
In effect, the average over equivalent strategies has cut off some of the
 original components, but has left the remaining ones
 (${\cal M}^{\,'}_{\odot,II,JJ}(\alpha)$ and ${\cal M}^{\,'}_{\odot,IJ,IJ}(\alpha)$)
unchanged.
\bigskip

Due to the blockform of (\ref{Modot}), any of the operators ${\cal
M}^{\,\,'}_\odot(\alpha)$ leaves two subspaces of ${\cal B}$
invariant: ${\cal W}$, the $d$-dimensional subspace spanned by the
basis elements $\{e_{II}|I=1,\dots d\}$ (containing ${\mathbf{1}}$
and $\tau$), and its $d(d-1)$-dimensional orthogonal complement
${\cal W}^\perp$, spanned by the basis elements
$\{e_{IJ}|I,J=1,\dots d,I\neq J\}$. Thus, it uniquely decompose
into the direct sum ${\cal M}^{\,\,'}_\odot(\alpha) = {\cal
R}(\alpha)\oplus S$, where ${\cal R}(\alpha)$ acts on ${\cal W}$,
and $S$ acts on ${\cal W}^\perp$. The components of these
operators are
\begin{eqnarray}
{\cal R}_{IJ}(\alpha)&=&\,\,\,{\cal M}^{\,'}_{\odot,II,JJ}(\alpha)\qquad\,\,\,\,\forall I,J\label{Rdef}\\
S_{IJ,KL}&=&
\left\{
\begin{array}{cl}
{\cal M}^{\,'}_{\odot,IJ,IJ}(\alpha)\quad&{\rm for\,\,} (IJ)=(KL),I\neq J, K\neq L\\
0\quad&{\rm for\,\,}(IJ)\neq(KL),I\neq J, K\neq L\label{Sdef}
\end{array}
\right.
\end{eqnarray}
Since the indices of $S$ are understood as pairs $IJ$ with ($I\neq J$),
the array $S_{IJ,KL}$ forms a diagonal $d(d-1)\times d(d-1)$ matrix.
As indicated, it is independent of $\alpha$ (because the operator ${\cal P}$ as defined in
(\ref{calP}) acts as a projection in ${\cal W}$ and annihilates ${\cal W}^\perp$).
Some algebra shows how our measures of knowledge may be expressed in terms of these
objects: Let
\be
R_{IJ}\,=\,{\cal R}_{IJ}(0)\,,
\label{R0def}
\ee
which is the $d\times d$ matrix made up be the $IIJJ$ components of (\ref{MB})
 when the $\alpha{\cal P}$-term is ignored, and
\be
E_{IJ}\,=\,1\qquad\forall I,J
\ee
reflecting the component structure of the $\alpha{\cal P}$-term in (\ref{MB}). Then
\begin{eqnarray}
{\rm det}\,{\cal M}^{\,'} &=& \frac{1}{d}\,\,{\rm det} R\,\,\,{\rm Tr}(R^{-1}E)\,\,{\rm det}\,S
\label{Mcalc}\\
{\rm Tr}({\cal M}^{\,'\,-1}) &=& {\rm Tr}(R^{-1})-\frac{{\rm Tr}(R^{-1}E R^{-1})}{{\rm Tr}(R^{-1}E)}
+ {\rm Tr}(S^{-1})\,.\label{Trcalc}
\end{eqnarray}
When computing these two quantities one may use the fact that they
are invariant under the replacement $R\rightarrow R+c E$ for any
constant $c$. The combination ${\rm det}R\times{\rm Tr}(R^{-1}E)$
may likewise be written as $\sum_{I,J}(-)^{I+J}{\rm det}_{IJ}R$,
where ${\rm det}_{IJ}R$ is the determinant of the matrix obtained
from $R$ by deleting the $I$-the row and the $J$-th column.
(We recall from linear algebra that ${\rm det}_{IJ}R =
{\rm det}R\,(R^{-1})_{JI}$). It thus follows that ${\rm
det}\,{\cal M}^{\,'}$ is a polynomial expression in $R_{IJ}$.
\bigskip
\bigskip

\noindent {\bf Comparison of efficiency for different states }
\medskip

\noindent In order to compare the efficiency of an improved strategy for different states
in the volume oriented approach, we note that, according to (6.6) and (6.8),
${\cal M}(\lambda \tau _1+(1-\lambda )\tau _2) \leq \lambda {\cal
M} (\tau _1)+(1-\lambda ){\cal M}(\tau _2)$ 
for $0\leq\lambda\leq 1$. Hence, as in (6.15) and (6.16), 
the Peierls-Bogoliubov inequality guarantees that
the strategy is more efficient for a state that is less mixed, i.e.
\begin{eqnarray}
{\rm det}_\odot\,{\cal M}^{\,'}_{\odot}(\alpha,\lambda \tau _1
+(1-\lambda )\tau _2)  &\leq& \lambda \,{\rm det}_\odot\,{\cal
M}^{\,'}_\odot(\alpha, \tau _1)\nonumber\\
&&+(1-\lambda ) \,{\rm det}_\odot\,{\cal
M}^{\,'}_\odot (\alpha, \tau _2).
\end{eqnarray}
Later on, when discussing particular strategies, we will concentrate 
on ``typical'' states, i.e. the tracial state,
which is maximally mixed, and states with some vanishing eigenvalues.
\bigskip
\bigskip

\noindent {\bf Strategy 1: Using mutually unbiased observables}
\medskip

\noindent From our result for the two-dimensional case we guess
that it is a good strategy to choose one observable in the
direction of $\tau$, e.g. $B_1 =\tau$. 
For simplicity, we assume that $\tau$ is
non-degenerate, its spectral projections thus being the
one-dimensional operators $e_{II}$. If $\tau$ is degenerate,
we slightly change it to some non-degenerate $\tilde{\tau}$ and
re-insert $\tau$ in the very end of the computation.
Following the spirit of Wootters and Fields
\cite{WoottersFields1989}, we seek to choose the other observables
$B_\beta$ ($\beta \geq 2$) such that all eigenbases are mutually unbiased. It is not known
whether for arbitrary dimensions $d$ such operators exist.
However, the averaging method as developed above provides a
strategy that comes close to this idea and is realizable in any
dimension. It requires just {\it one} other observable, $B_2$,
satisfying
\be
{\rm Tr}(P_{1 a} P_{2 a'}) \,=\, \frac{1}{d}
\quad\forall a\in{\rm Sp}(B_1) {\rm \,\,and\,\,} a'\in{\rm Sp}(B_2).
\label{mu}
\ee
This may also be written as
\be
P_{2 a,II} \,=\, \frac{1}{d}\quad\forall a\in{\rm Sp}(B_2)
\ee
and implies $(\tau|P_{2a})=2/d\,\,\forall a\in{\rm Sp}(B_2)$.
To these two observables we apply the strategy improving mechanism (\ref{Modot}).
{\it If} however there exists a large enough family of mutually unbiased bases,
as in the explicit example given in \cite{WoottersFields1989},
then all components $P_{\beta a,IJ}$ of $P_{\beta a}$ coincide up to phase factors, 
and we expect the strategy based
on these to be equivalent to the one we will now analyze.
(In the two-dimensional case, this corresponds to the fact that
we can either measure in two fixed orthogonal directions -- as has explicitly been
 worked out in the preceding section --, or alternatively in all directions
orthogonal to ${\vec u}$. In this case the averaging method does
not lead to anything new).
\bigskip

So let us start with $B_1=\tau$ and $B_2$ satisfying (\ref{mu}). We leave $n_1$ and $n_2$
unspecified for the moment. With (\ref{MB}), the contributions to (\ref{Modot}) for $\alpha=0$
are as follows:
\begin{eqnarray}
n_1\,{\cal Q}_{II,JJ}(B_1)&=&n_1\,\sum_{a\in{\rm Sp}(B_1)}\frac{P_{1a,II}\,P_{1a,JJ}^{\,\,*}}{(\tau|P_{1a})}
\,\,=\,\,\frac{n_1}{2\,\tau_{II}}\,\delta_{IJ}\quad\forall I,J\label{cont1}\\
n_1\,{\cal Q}_{IJ,IJ}(B_1)&=&n_1\,\sum_{a\in{\rm Sp}(B_1)}\frac{|P_{1a,IJ}|^2}{(\tau|P_{1a})}
\,\,=\,\,0\quad {\rm for\,\,}I\neq J\label{cont2}\\
n_2\,{\cal Q}_{II,JJ}(B_2)&=&n_2\,\sum_{a\in{\rm Sp}(B_2)}\frac{P_{2a,II}\,P_{2a,JJ}^{\,\,*}}{(\tau|P_{2a})}
\,\,=\,\,\frac{n_2}{2}\quad\forall I,J\label{cont3}\\
n_2\,{\cal Q}_{IJ,IJ}(B_2)&=&n_2\,\sum_{a\in{\rm Sp}(B_2)}\frac{|P_{2a,IJ}|^2}{(\tau|P_{2a})}
\,\,=\,\,\frac{n_2}{2}\quad {\rm for\,\,}I\neq J.\label{cont4}
\end{eqnarray}
Adding (\ref{cont1})+(\ref{cont3}) and (\ref{cont2})+(\ref{cont4}) gives all non-zero components of ${\cal M}^{\,\,'}_\odot(0)$.
The only nonzero components of the operators $R$ and $S$ as
introduced in (\ref{Rdef})--(\ref{R0def}) are thus given by
\begin{eqnarray}
R_{IJ}&=&
\frac{n_1}{2\,\tau_{II}}\,\delta_{IJ} + \frac{n_2}{2}\quad\forall I,J\\
S_{IJ,IJ}&=&\frac{n_2}{2}\quad {\rm for\,\,}I\neq J.
\end{eqnarray}
Using (\ref{Mcalc}), our final result for the volume oriented approach reads
\be
{\rm det}{\cal M}^{\,\,'} \,=\,
\frac{1}{d}\,\left(\frac{n_1}{2}\right)^{d-1}\,\left(\frac{n_2}{2}\right)^{d(d-1)}\,{\rm det}(\tau^{-1})\,.
\label{finM}
\ee
For given $n=n_1+n_2$, the best of all these strategies is characterized by
$n_1 n_2^d={\rm max}$, which leads to
\be
n_1=\frac{n}{d+1}\qquad{\rm and}\qquad n_2=\frac{d\,n}{d+1},
\label{str1}
\ee
hence
\be
({\rm det}{\cal M}^{\,\,'})_{\rm max} \,=\,
d^{d^2-d-1} \,\left(\frac{n}{2(d+1)}\right)^{d^2-1}\,{\rm det}(\tau^{-1})\,.
\label{detmax1}
\ee
If $d=2$, this coincides with the value (\ref{detmax}) for the best two-dimensional
 (volume oriented) strategy.
However, as we shall see, in higher dimensions there are states $\tau$ for which one
 can do better.
Analogously, using (\ref{Trcalc}), we find for the distance oriented approach
\be
{\rm Tr}({\cal M}^{\,'\,-1})\,=\,\frac{2\,}{n_1}\,\left(1-{\rm Tr}(\tau^2)\right)
+ \frac{2\,d(d-1)}{n_2}\,.
\ee
For given $n=n_1+n_2$, the best of all these strategies are characterized by
\be
n_1=\frac{n}{1+\sqrt{\frac{d(d-1)}{1-{\rm Tr}(\tau^2)}}}
\qquad{\rm and}\qquad
n_2=\frac{n}{1+\sqrt{\frac{1-{\rm Tr}(\tau^2)}{d(d-1)}}}\,,
\label{str2}
\ee
hence
\be
{\rm Tr}({\cal M}^{\,'\,-1})_{\rm min}\,=\,
\frac{2}{n}\left(\sqrt{1-{\rm Tr}(\tau^2)}+\sqrt{d(d-1)}\,\right)^2.
\label{Trmin1}
\ee
If $d=2$, this coincides with the value (\ref{Trmin}) for the best two-dimensional
 (distance oriented) strategy.
Whether one can do better in higher dimensions is an open question.
\bigskip

Summarizing, the strategies specified by (\ref{str1}) and (\ref{str2})
 are in a sense the natural generalizations
from the two-dimensional case, their effectiveness being quantified by
(\ref{detmax1}) and (\ref{Trmin1}).
\bigskip
\bigskip

\noindent {\bf Strategy 2: Using matrix units}
\medskip

\noindent We will now -- for even dimensions -- construct a
different strategy that sometimes works better in the volume
oriented approach. From the two-dimensional situation we have
learned the following: The uncertainties (in both the volume and
the distance oriented approach) are smaller when the unknown state
is less mixed. As in the strategy constructed above, we choose one
observable, $B_1$, coinciding with $\tau$. The other observables
should give as much new information as possible, therefore should
be sufficiently independent of $\tau $. They are maximally
independent if they are mutually unbiased. However, then the
uncertainties tend to be large. Therefore two effects are
competing, and we have observed that in two dimensions the
independency is the dominating effect. In higher dimensions, a
convenient basis of ${\cal B}$ is given by the matrix units
(\ref{matrixunits}), constructed out of an eigenbasis of $\tau$.
Since these operators are not positive (not even hermitian) and
therefore do
 not correspond to observables,
we resort to the $d(d-1)$ projections defined by ($I<J$)
\be
P_{IJ}^{\,\pm} \,=\,\frac{1}{2}\left(e_{II}\pm e_{IJ} \pm e_{JI} + e_{JJ}\right)\,.\label{P}
\ee
Our goal is to construct the rest of our observables out of these operators.
As before, we understand that the average procedure (\ref{Modot}) has been performed.
In effect this just means to
take into account only the components of ${\cal M}^{\,\,'}_{\odot,IJ,KL}(0)$
relevant for $R_{IJ}$ and $S_{IJ,IJ}$ as defined in
(\ref{Rdef})--(\ref{R0def}). Any $P_{KL}^{\,\pm}$ ($K<L$) appearing as spectral projection
of an observable $B_\beta$ gives the contributions
\begin{eqnarray}
n_\beta\,\frac{P^{\,\pm}_{KL,II}\,P_{KL,JJ}^{\,\pm\,*}}{(\tau|P_{KL}^{\,\pm})}&
=&\frac{n_\beta}{4}\,\frac{(\delta_{IK}+\delta_{IL})(\delta_{JK}+\delta_{JL})}
{\tau_{KK}+\tau_{LL}}
\quad\forall I,J\label{PKL1}\\
n_\beta\,\frac{|P^{\,\pm}_{KL,IJ}|^2}{(\tau|P_{KL}^{\,\pm})}&
=&\frac{n_\beta}{4}\,\frac{\delta_{IK}\delta_{JL}+\delta_{JK}\delta_{IL}}{\tau_{KK}+\tau_{LL}}
\quad {\rm for\,\,}I\neq J\label{PKL2}
\end{eqnarray}
to $R_{IJ}$ and $S_{IJ,IJ}$, respectively.
These expressions have to be summed up for all projectors involved.
The contributions from $B_1$ are identical with (\ref{cont1})--(\ref{cont2}).
\bigskip

Let us now show how the projections (\ref{P}) may be used to define suitable observables.
The idea is to group these operators into $d-1$ subfamilies, each containing $d$ elements,
in order to construct $d-1$ observables in addition to $B_1$.
We will restrict ourselves to even $d$ and define $B_2$ to have the
spectral projections (the eigenvalues being irrelevant as long as each observable is non-degenerate)
\be
P_{12}^{\,+},\,P_{12}^{\,-},\,P_{34}^{\,+},\,P_{34}^{\,-},\,\dots P_{(d-1)\,d}^{\,\,+},\,P_{(d-1)\,d}^{\,\,-}\,.
\ee
This may be abbreviated in terms of the partition
\be
B_2\,\longleftrightarrow\,(1,2)(3,4)\dots(d-1,d)
\ee
of $(1,2,\dots,d)$. The remaining observables are obtained by appropriately
permuting certain numbers in the above partition, such that any pair never occurs twice.
This is possible in any even dimension and can best be explained in an example:
 For illustration we choose $d=6$ and define
\begin{eqnarray}
B_2&\longleftrightarrow&(1,2)(3,4)(5,6)\nonumber\\
B_3&\longleftrightarrow&(1,3)(2,5)(4,6)\nonumber\\
B_4&\longleftrightarrow&(1,5)(3,6)(2,4)\\
B_5&\longleftrightarrow&(1,6)(5,4)(3,2)\nonumber\\
B_6&\longleftrightarrow&(1,4)(6,2)(3,5)\nonumber
\end{eqnarray}
The underlying general procedure is the following: One number in every pair is
 moving to the right,
one to the left as long as it is possible, then it is reflected.
In this way every number corresponds to a line, and every line crosses every other line
exactly once. For $d>4$ there are other possible permutation schemes
(which should all be taken into account when the best of these strategies is
 to be determined).
A strategy is fixed by giving any observable $B_\beta$ ($\beta=1,\dots d$)
 a weight $n_\beta$,
the number of measurements reserved for the family $B^{\,'}_\beta(\varphi)$, such that
$\sum_{\beta=1}^dn_\beta=n$.
\bigskip

In order to write down the operators $R$ and $S$ for this type of strategy, we note that,
for given $K,L$ ($K\neq L$), either the pair $P_{KL}^{\,\pm}$ or the pair $P_{LK}^{\,\pm}$
occurs in some $B_\beta$ ($\beta\geq 2$).
Let us denote this $\beta$ by $\beta(K,L)$. Using this notation, we
sum up (\ref{PKL1}) and (\ref{cont1}) for $R$, (\ref{PKL2}) and (\ref{cont2}) for $S$,
to obtain
\begin{eqnarray}
R_{IJ}&=&\delta_{IJ}
\left(\frac{n_1}{2\,\tau_{II}} + \sum_{K\neq I}\frac{n_{\beta(I,K)}}{2\,(\tau_{II}+\tau_{KK})}
\right)+
(1-\delta_{IJ})\,\frac{n_{\beta(I,J)}}{2\,(\tau_{II}+\tau_{JJ})}\label{Rcal}\\
S_{IJ,IJ}&=&\frac{n_{\beta(I,J)}}{2\,(\tau_{II}+\tau_{JJ})}
\quad {\rm for\,\,}I\neq J.\label{SS}
\end{eqnarray}
The explicit evaluation of (\ref{Mcalc}) and (\ref{Trcalc}) for a general strategy of this
type and comparison with our previous results (\ref{detmax1}) and (\ref{Trmin1})
is not an easy task. We will therefore confine ourselves to a family of examples:
Let $d\geq 4$, $\tau_{11}=\tau_{22}=a/2$ and $\tau_{33}=\dots=\tau_{dd}=(1-a)/(d-2)$,
and set $n_\beta=n'$ for all $\beta=2\dots d$ (i.e. $n_{\beta(I,J)}=n'$ for all $I\neq J$).
For small $a$, the combination $(\tau_{11}+\tau_{22})^{-1}$ is large. This blows up the
determinant of ${\cal M}$: We find $R_{11}= R_{22}= n_1/a+O(1)$ and
$R_{12}= R_{21}=S_{1212}=S_{2121}= n'/(2a)+O(1)$, whereas all other components
are finite for $a\rightarrow 0$. The application of (\ref{Mcalc}) to (\ref{Rcal})--(\ref{SS})
exhibits the behaviour
\be
{\rm det}{\cal M}^{\,\,'}\,\sim\,O(a^{-4})\qquad{\rm for\,\, small\,\,}a.
\ee
This may be compared with (\ref{detmax1}) which -- for the same $\tau$ --
diverges only as $O(a^{-2})$.
Hence, for given even dimension $d\geq 4$, there is always an unknown state $\tau$
(defined by sufficiently small $a$) such that a
strategy of type 2 is better than strategy 1 in the volume oriented approach.
For the distance oriented approach, there is no such difference in the
scaling behaviour for $a\rightarrow 0$.
\bigskip

For $a=2/d$, we obtain the tracial state $\tau = d^{-1}\,{\mathbf{1}}$, i.e.
$\tau_{II}=d^{-1}$ for all $I$. In this case we can be more explicit, and we obtain
\begin{eqnarray}
{\rm det}\,{\cal M}^{\,'} &=& \left(\frac{d}{4}\right)^{d^2-1}
\left([2n_1+n'(d-2)]\,n'^d\right)^{d-1}
\label{McalcMU}\\
{\rm Tr}({\cal M}^{\,'\,-1}) &=& 4(d-1)\left(\frac{1}{[2n_1+n'(d-2)]\,d}+\frac{1}{n'}\right).
\label{TrcalcMU}
\end{eqnarray}
Interestingly, if $d\geq 4$, both expressions become optimized if $n_1=0$, i.e. $n'=n/(d-1)$.
Hence, the best values for this class of strategies for the tracial state are given by
\begin{eqnarray}
({\rm det}\,{\cal M}^{\,'})_{\rm max} &=&
\left(\frac{d}{4}\right)^{d^2-1}
\left((d-2)\left(\frac{n}{d-1}\right)^{d+1}\right)^{d-1}
\label{McalcMUmax}\\ {\rm Tr}({\cal M}^{\,'\,-1})_{\rm min} &=&
\frac{4(d-1)^4}{n (d-2)d}\,. \label{TrcalcMUmin}
\end{eqnarray}
The volume oriented strategy (\ref{McalcMUmax}) gets beaten by
(\ref{detmax1}), because -- for the tracial state -- (\ref{detmax1})$\geq$(\ref{McalcMUmax}) for all $d$.
Asymptotically for large $d$, (\ref{detmax1}) exceeds
(\ref{McalcMUmax}) by a factor of leading order $2^{d^2}$.
Similarly, the distance oriented strategy (\ref{TrcalcMUmin}) is worse than 
(\ref{Trmin1}), because, for large $d$, (\ref{TrcalcMUmin}) is twice as large as (\ref{Trmin1})
for the tracial state.
\bigskip

It is easy to show that the last feature remains true for general states
if $d\geq 6$: Using the estimates
\begin{eqnarray}
&&{\rm Tr}(S^{-1})=\sum
_{I\neq J} S_{IJ,IJ}^{-1}=\sum _{I\neq J}
\frac{2(\tau_{II}+\tau_{JJ})}{n_{\beta(I,J)}}
 =\sum _{\beta
}\frac{2}{n_{\beta }}\geq \frac {4(d-1)^2}{n-n_1}\\
&&{\rm
Tr}(R^{-1})-\frac{{\rm Tr}(R^{-1}E R^{-1})}{{\rm Tr}(R^{-1}E)}\geq 0
\end{eqnarray}
together with (\ref{Trcalc}), we find
\be
{\rm Tr}({\cal M}^{\,'\,-1})\geq \frac {4(d-1)^2}{n}.
\ee
From this it follows that also for general states in $d\geq 6$ our strategy of type 2 
cannot beat (\ref{Trmin1}).
\bigskip

Summarizing, for even dimensions $\geq 4$, there are states
$\tau$ for which the strategy (\ref{detmax1}) based on mutually unbiased
observables is not optimal when evaluated in the volume oriented approach.
On the other hand, in the distance oriented approach, we cannot offer 
a strategy better than (\ref{Trmin1}).
\bigskip
\bigskip

\noindent {\bf Remarks on infinite dimensions}
\medskip

\noindent The
results achieved in this paper suggest that the number of
measurements necessary in order to arrive at an estimate of the
unknown state $\tau$ with an uncertainty of the order $\epsilon$
increases like $d^2$ with increasing dimension $d$ of the Hilbert
space ${\cal H}$. This may be seen in both approaches we
discussed: Identifying $\epsilon^2$ with ${\rm Tr}({\cal M}^{-1})$
in the distance oriented approach, (\ref{Trmin1}) implies
$\epsilon^2\approx 2d^2/n$, hence $n\sim(d/\epsilon)^2$. The
analogous situation for the volume oriented approach is roughly
modeled by identifying $({\rm det}{\cal M})^{-1/2}$ with the
volume of a sphere of radius $\epsilon$ in $(d^2-1)$-dimensional
Euclidean space. Using Stirling's formula, the latter is for large
$d$ given by
\be
{\cal V}_d\,\approx\,\frac{\epsilon^{d^2-1}}{\sqrt{(d^2-1)\pi}}
\left(\frac{2e\pi}{d^2-1}\right)^{(d^2-1)/2}.
\ee
With (\ref{detmax1}) -- the strategy based on mutually unbiased observables -- 
and fixed ${\rm det}(\tau^{-1})$, this gives $n\sim(d/\epsilon)^2$
as in the distance oriented approach.
This behaviour is confirmed by the strategies of type 2 (using matrix units)
which were found to be better for certain states.
Since we need an additional number of measurements to get a first rough
estimate of $\tau$, the formula $n\sim(d/\epsilon)^2$ is to be understood as the
leading asymptotic behaviour of $n$ as $\epsilon$ approaches $0$.
\bigskip

If this behaviour is true for the best strategies possible, it has dramatic consequences
for the infinite dimensional case:
At first glance, it would altogether be impossible to determine $\tau$ with some
(given) uncertainty $\epsilon$. However, in infinite dimensions we may
decompose the Hilbert space as ${\cal H}=P_d({\cal H})\oplus P_d({\cal H})^\perp$,
where $P_d$ is some finite ($d$-)dimensional hermitean projection, and
measure $P_d$ in a number of copies of our quantum system.
Starting with $d=1$, we choose a one-dimensional hermitean projection $P_1$.
Whenever the measurement outcome is $0$, i.e. corresponds to $P_d({\cal H})^\perp$,
we redefine $d_{\rm new}=d+1$, choose some new decomposition such that
$P_{d_{\rm new}}\geq P_d$, and proceed analogously.
During this process, the probability for $0$ to occur in a further measurement,
given by $1-{\rm Tr}(\tau P_d)$, drops down to zero as $d$ increases.
In other words, the measurement data become increasingly consistent with
the expectation that $\tau$ is a density matrix in $P_d({\cal H})$.
If $\rho_{\rm ex}$ is the expected state, the uncertainty $\epsilon$ about $\tau$ is
given by $\epsilon^2\approx{\rm Tr}((\tau-\rho_{\rm ex})^2)$.
In terms of an appropriate block matrix notation we have
\be
\rho_{\rm ex} =\left(\begin{array}{cc}
\rho_d & 0\\
0 & 0
\end{array}\right)
\qquad\tau = \left(\begin{array}{cc}
\tau_d & \nu_d\\
\nu_d^\dagger & \tau_\infty
\end{array}\right),
\ee
so that
${\rm Tr}((\tau-\rho_{\rm ex})^2)={\rm Tr}((\tau_d-\rho_d)^2)
+2\,{\rm Tr}(\nu_d^\dagger\nu_d) + {\rm Tr}(\tau_\infty^2)$.
For given $\epsilon_0>0$, there is a (finite) dimension $d_{\rm eff}$ and a (finite)
 number $n_0$ of
measurements necessary to make
sure that $2\,{\rm Tr}(\nu_{d_{\rm eff}}^\dagger\nu_{d_{\rm eff}}) + {\rm Tr}(\tau_\infty^2)\,\lsim\,\epsilon_0^2\,$.
The numbers $d_{\rm eff}$ and $n_0$ will depend on $\tau$ and on the sequence of projections $P_1,P_2,\dots$ chosen.
Once having reached this point, we proceed {\it as if}
$\tau$ acts entirely in the subspace $P_{d_{\rm eff}}({\cal H})$. (Technically,
 we measure observables of the type
$A\oplus({\mathbf{1}}-P_{d_{\rm eff}})$, where $A$ acts in $P_{d_{\rm eff}}({\cal H})$,
 and ignore further outcomes that
belong to the remaining infinite dimensional subspace
$P_{d_{\rm eff}}({\cal H})^\perp$). Next, given $\epsilon_1$, we need a further
number $n_1\sim (d_{\rm eff}/\epsilon_1)^2$ of measurements to arrive at a final
 estimate
$\rho_{\rm fin}$ in $P_{d_{\rm eff}}({\cal H})$ such that
${\rm Tr}((\tau_{d_{\rm eff}}-\rho_{\rm fin})^2)\,\lsim\,\epsilon_1^2\,$.
Hence, after $n=n_0+n_1$ measurements, the total uncertainty is
of the order $\epsilon=(\epsilon_0^2+\epsilon_1^2)^{1/2}$.
This procedure
enables one to determine an unknown state
to any desired degree of security even if it lives in an infinite dimensional Hilbert
space. (This is of course not an optimal strategy.
A more efficient method is e.g. to combine the two parts of the procedure
and to measure observables of the type $A\oplus({\mathbf{1}}-P_d)$ from the outset).
\bigskip

The apparent contradiction of this result with the behaviour $n\sim(d/\epsilon)^2$
in the case of large but finite dimension $d$ is clarified by noting that
the number $n_0$ may be very large: Suppose some sequence of projections
$P_{d+1}=P_d+|e_{d+1}\rangle\langle e_{d+1}|$ has been fixed
($e_I$ denoting an orthonormal basis of ${\cal H}$,
the starting point being $P_1=|e_1\rangle\langle e_1|$),
and suppose that $\tau = |e_D\rangle\langle e_D|$ for some $D$ (that may be very large).
In this case it takes $D$ measurements until a non-zero outcome is possible.
Similar scenarios are possible for any $\tau$: Given an arbitrary number
$N$, then (with some portion of bad luck) it is always possible to
adjust the sequence of projections such that $n_0>N$.
Hence, there exists no general upper bound for $n_0$ (and thus for $n$).
This feature is not present in the finite dimensional case.
The behaviour $n\rightarrow\infty$ as obtained by
letting $d\rightarrow\infty$ in the formula $n\sim(d/\epsilon)^2$
must be understood in this sense.
\bigskip

A problem still persists with our approach. It stems from the fact that we have invoked a Gaussian approximation.
For finite dimension $d$ we infer from (\ref{appr}) and (\ref{app2}) that this
approximation is reliable if $\epsilon\!\parallel\!\tau^{-1}\!\parallel\,\ll 1$.
In other words: if $\epsilon$ is chosen too large, our formalism will fail to
reproduce the unknown state with the promised accuracy. 
As a consequence, we must have $n\gg d^2\!\parallel\!\tau^{-1}\!\parallel$,
which means that a smaller number of measurements will not lead to a 
reasonable result. This introduces an additional dependence on the dimension
into the state determination problem:
Since $\parallel\!\tau^{-1}\!\parallel\ge d$, we have 
$\epsilon\ll d^{-1}$ and $n\gg d^3$. However, in large dimensions, typical 
density matrices tend to have even larger $\parallel\!\tau^{-1}\!\parallel$.
In the infinite dimensional case, $\parallel\!\tau^{-1}\!\parallel$ is no longer finite.
Even when reducing the problem to an effectively finite dimensional one, as sketched above,
we can expect the density matrix $\tau_{d_{\rm eff}}$ 
to have a very large (if not infinite) value of $\parallel\!\tau_{d_{\rm eff}}^{-1}\!\parallel$.
This in turn requires the choice of a correspondingly small $\epsilon$
and blows up $n$. A partial cure of this dilemma is to modify the determination of 
$P_{d_{\rm eff}}({\cal H})$ so as to statistically test 
any redefinition $d_{\rm new}=d+1$ whether a large enough portion of $\tau$ is
gained, and undo it otherwise. Thus, the small eigenvalues of 
$\tau$ may be kept in $P_{d_{\rm eff}}({\cal H})^\perp$, and only the large ones are taken into account.
In effect, we expect such a procedure to reduce the number of measurements
necessary.

\end{document}